\documentclass[twocolumn]{article}

\usepackage{amsmath}
\usepackage{booktabs}

\usepackage{graphicx,natbib}
\usepackage{authblk}
\usepackage{sectsty}
\sectionfont{\normalfont\large\bf}
\subsectionfont{\normalfont\it}

\usepackage{mathrsfs}

\def\unit#1{\,{\rm#1}}

\def\Msun{M_\odot}
\def\Mearth{M_\oplus}

\def\kg{\unit{kg}}

\def\kelvin{\unit{K}}
\def\metre{\unit{m}}
\def\sec{\unit{s}}

\def\joule{\unit{J}}
\def\photon{\unit{photons}}

\def\volt{\unit{V}}
\def\ohm{\unit{\Omega}}
\def\cvolt{\unit{V\!_{90}}}
\def\cohm{\unit{\Omega_{90}}}
\def\watt{\unit{W}}
\def\Hz{\unit{Hz}}

\def\THz{\unit{THz}}
\def\jansky{\unit{Jy}}
\def\ABmag{\unit{AB}}

\def\microsec{\unit{microsec}}

\def\au{\unit{au}}

\def\Gyr{\unit{Gyr}}

\def\pc{\unit{pc}}

\def\nue{\nu_e}
\def\nup{\nu_p}

\def\als{\bar a}
\def\Rls{\bar R}
\def\Dls{\bar D}
\def\vc{\beta}
\def\Mgs{\mathscr{M}}
\def\Egs{\mathscr{E}}
\def\hgw{h_{\rm GW}}
\def\lgdist{\bar R_{\rm\scriptscriptstyle M31}}

\def\up#1{\raisebox{0.5ex}{#1}}

\begin{document} 

\title{Astronomy and the new SI}

\author[1]{Prasenjit Saha}

\affil[1]{\small{\em Physik-Institut, University of Zurich,
  Winterthurerstrasse~190, 8057 Zurich, Switzerland}}

\setbox0=\vbox{\hsize=6.5truein \hbox to \hsize{\hss\bf
    Abstract\hss} \smallskip \noindent In 2019 the International
  System of units (SI) conceptually re-invented itself.  This was
  necessary because quantum-electronic devices had become so precise
  that the old SI could no longer calibrate them.  The new system
  defines values of fundamental constants (including $c,h,k,e$ but not
  $G$) and allows units to be realized from the defined constants
  through any applicable equation of physics.  In this new and more
  abstract SI, units can take on new guises --- for example, the
  kilogram is at present best implemented as a derived electrical
  unit.  Relevant to astronomy, however, is that several formerly
  non-SI units, such as electron-volts, light-seconds, and what we may
  call ``gravity seconds'' $GM/c^3$, can now be interpreted not as
  themselves units, but as shorthand for volts and seconds being used
  with particular equations of physics.  Moreover, the classical
  astronomical units have exact and rather convenient equivalents in
  the new SI: zero AB magnitude amounts to $\simeq5\times10^{10}$
  photons $\rm m^{-2}\,s^{-1}$ per logarithmic frequency or wavelength
  interval, $\rm 1\,au\simeq 500$~light-seconds, $\rm 1\,pc\simeq
  10^8$~light-seconds, while a solar mass $\simeq5$~gravity-seconds.
  As a result, the unit conversions ubiquitous in astrophysics can now
  be eliminated, without introducing other problems, as the old-style
  SI would have done.  We review a variety of astrophysical processes
  illustrating the simplifications possible with the new-style SI,
  with special attention to gravitational dynamics, where care is
  needed to avoid propagating the uncertainty in $G$.  Well-known
  systems (GPS satellites, GW170817, and the M87 black hole) are used
  as examples wherever possible. \par \medskip
   \par}

\date{\box0}

\maketitle

\section{Introduction}

While the whole point of units is to stay at fixed values, definitions
of units do in practice change from time to time, in response to
scientific developments.  From 1990 the SI (Syst\`eme International
d'unit\'es) faced a rebellion of electrical units.  The development of
Josephson junctions beginning in the 1960s provided voltage as
precisely $h/(2e)$ times a frequency.  The discovery of the quantum
Hall effect in 1980 provided a standard resistance $e^2/h$.  From
these two processes a new conventional volt $\!\cvolt$ and
conventional ohm $\!\cohm$ emerged, which were more precise than the
SI-prescribed standards involving force between current-carrying
wires.  Not only that, the derived electrical unit
\begin{equation}
\cvolt^2 \cohm\!\!^{-1} \metre^{-2} \sec^3
\end{equation}
was equal to a kilogram but more precise than the SI kilogram.  The
presence of an alternative standard that was more precise threatened
to make the SI irrelevant.  Faced with this crisis, the SI began a
long process of reinventing itself \citep[see][]{2001TaylorMohr}
leading to the reforms of 2018/2019.  In the new SI \citep{bipm2019}
only the second is specified by a specific physical process (a
spectral line in Cs).  Other units are defined implicitly, as whatever
makes the speed of light come out as $299792458 \metre\sec^{-1}$,
Planck's constant come out as $6.626070040\times10^{-34} \joule\sec$,
and so on. (Table~\ref{table:const} in the Appendix summarizes.)  Any
equation of physics may be used for realisation of units.  Thus, if
one wishes to interpret the kilogram as a derived electrical unit,
there is no longer a conflict with the SI.  A nice toy example where
this happens is the LEGO watt balance \citep{2015AmJPh..83..913C}
which measures mass as electrical watts times $\!\sec^3\metre^{-2}$.

Astronomers cannot be said to have rebelled against the SI, because
they never really joined.  In the 19th century astronomers set up
units standardized from the sky rather than in the laboratory.
Ancient stellar magnitudes were formalized as a logarithmic scale for
brightness \citep[see][for this
  history]{1968ASPL...10..145J,2013JHA....44...47S}.  The Sun became
the prototype mass.  Most interesting was the astronomical unit of
length \citep[for the history, see][]{1970SvA....13..712S}.
Anticipating the explicit-constants style of the new SI, the au was
defined such that the formal angular velocity $\sqrt{G\Msun/\au^3}$
equals exactly 0.01720209895 radians per day (which is close to $2\pi$
per sidereal year).

Modern practice has continued with the classical units of length,
mass, and brightness, but calibrated them against the SI.
\begin{itemize}
\item The astronomical unit of length is now defined as
  $1\au=149\,597\,870\,700\metre$.  A parsec is the distance at which
  $1\au$ subtends one arc-second, hence now also defined as a fixed
  (albeit irrational) number of metres.  This calibration of
  astronomical distances to the SI, retiring the 19th-century
  definition of the au, was adopted quite recently, in
  2012.\footnote{Resolution B2 {\em on the re-definition of the
      astronomical unit of length,} adopted at the IAU General
    Assembly (2012).}
\item The Sun's mass times the gravitational constant (known as the
  solar mass parameter) is nowadays stated in SI
  units.\footnote{Resolution B3 {\em on recommended nominal conversion
      constants for selected solar and planetary properties,} adopted
    at the IAU General Assembly (2015).}  It is measured as
  \begin{equation}\label{eq:GMsun}
  G\Msun=1.3271244\times10^{20}\metre^3\sec^{-2}
  \end{equation}
  in Newtonian dynamics. Two more significant digits are available if
  general-relativistic time dilation is included, which we will
  discuss later (Section~\ref{subsec:timedil}).
\item The AB-magnitude scale \citep{1983ApJ...266..713O} sets a
  spectral flux density of a jansky
  (that is, $10^{-26}\watt \metre^{-2} \Hz^{-1}$) as $\!\ABmag=8.90$.
  This style is not unique to astronomy --- in the moment-magnitude
  scale in seismology \citep{1979JGR....84.2348H} a zero-magnitude
  earthquake corresponds to a gigajoule (or rather $10^{9.05}\joule$),
  and two magnitudes correspond to a factor of a thousand in energy.
\end{itemize}
Along with these SI-calibrated classical units, astronomers commonly
use several named units that are SI units times a simple
multiplicative factor.  Some of them are universally understood
(hours, minutes, also degrees, arc-minutes and arc-seconds) and
recognized in Table~8 of the SI brochure as useful alongside SI units.
Others (ergs, gauss, and \AA) are legacies of pre-SI conventions.

But there remains one important unsolved problem in calibrating
conventional astronomical units against the SI, and again it has to do
with the kilogram.  In the laboratory, mass is a measure of inertia,
but in astrophysics, mass is observable from the gravitational field
it produces.  Whether the mass is the Sun, or an asteroid less than
100~km across and $\sim10^{-12}\Msun$
\citep[e.g.,][]{2014A&A...565A..56G}, the astrophysical observable is
not $M$ but $GM$.  Unfortunately, $G$ is known only to $10^{-4}$
\citep{2018Natur.560..582L}, and while there are some creative new
ideas for measuring $G$
\citep{2005MNRAS.356..587C,2014Natur.510..518R}, there is no near-term
prospect of more than four significant digits.  The error propagates
to any astrophysical mass expressed in kilograms, and in particular to
$\Msun=1.988(2)\times10^{30}\kg$.  This much uncertainty in the mass
of the Sun would be fatal for precision applications like spacecraft
dynamics.  Hence, kilograms cannot be used to express gravitating
masses in precision applications.  With the kilogram unusable, it is
not surprising that astronomers have little appetite for changing any
of their conventions to SI units.

The new SI, however, changes the situation.  There are no new units in
the new SI, nor do any values change significantly, instead the reform
is conceptual.  In particular, the kilogram is now defined implicitly.
This suggests leaving astrophysical masses in kilograms implicit
(pending future developments in the measurement of $G$) and working
with the mass parameter in SI units.  Of course, in solar-system
dynamics, this is already done, only $\!\metre^3\sec^{-2}$ seems too
un-memorable for general use.  But that problem is easily remedied, by
using $GM/c^2$ in metres, or $GM/c^3$ in seconds.  Basically, one
could work with the formal Schwarzschild radius in place of the mass.

This paper will develop some ideas like the above.  The aim is not
some kind of formal compliance with the SI, but to suggest some
formulations that (a)~have a precise meaning in the new SI,
(b)~simplify formulas and help understand astrophysical processes
better, and (c)~would be reasonably easy to convert to.  This paper
will focus on the au, pc, $\Msun$, and optical magnitudes.  If those
get converted, units like the \AA\ and gauss can be trivially
replaced, and do not need discussion here.  Section~\ref{sec:defs}
discusses SI formulations that are equivalent to the classical
astronomical units, but have somewhat different interpretations.
Section~\ref{sec:examples} is devoted to example applications, where
we see that replacing the au, pc, solar mass, optical magnitudes with
SI equivalents is quite convenient and can provide insight into
diverse astrophysical phenomena.

\section{Length, mass, and brightness units}\label{sec:defs}

Astronomical distances are often stated in light seconds.  In topics
where general relativity plays a role, $GM/c^3$ in seconds may appear
as a surrogate for mass.  AB magnitudes are equivalent to photon flux
per logarithmic wavelength interval \citep[see Eq.~1
  from][]{2012ApJ...750...99T}.  Let us discuss these in turn.

\subsection{Light-seconds}

The light-second is a common and useful informal unit, and has
conveniently round conversions:
\begin{equation}\label{eq:approx-lsec}
\begin{aligned}
1 \au  &\longrightarrow 5.0\times10^2 \sec \\
1 \pc  &\longrightarrow 1.0\times10^8 \sec \,.
\end{aligned}
\end{equation}
Table~\ref{table:conv} in the Appendix gives precise conversions.

To use light-seconds more than informally (that is, in formulas and
computer programs), we need to decide whether a light second is a
length or a time.  That is, a light-second could be a synonym for
$299792458\metre$, {\em or} it could be this length divided by $c$.
In this paper, the latter interpretation will be used.  That is, a
light-second will be taken as a normal second measurable by a clock,
just being used to measure a length divided by $c$.

An analogous situation applies to the electronvolt.  The list of
useful non-SI units in Table~8 of the SI brochure gives the
electronvolt as $1.602176634\times10^{-19}\joule$.  But the
electronvolt is also used to measure mass or even frequency.  Hence,
it is more useful to understand the electronvolt as just a volt, with
the `electron' label denoting that, according to context, one is
measuring energy divided by the electron charge $e$, or mass times
$c^2/e$, or frequency times $h/e$.  Formerly, such a multiplication
depended on the experimentally-determined value of $e$, but not any
more, because in the new SI $e$ is a defined constant.

Later in this paper, we will write several distances in light seconds.
For this, let us fix the notation
\begin{equation}\label{eq:barls}
\begin{pmatrix}
\als \\[5pt] \Rls \\[5pt] \Dls
\end{pmatrix}
\equiv
\begin{pmatrix}
a/c \\[5pt] R/c \\[5pt] D/c
\end{pmatrix}
\end{equation}
to express lengths as times.

\subsection{Gravity seconds}

We could choose $GM/c^2$ in metres as an easier variant of the mass
parameter, but $GM/c^3$ in seconds blends somewhat better with
light-seconds.  There is no standard term for measuring
\begin{equation}
\Mgs \equiv \frac{GM}{c^3}
\end{equation}
in seconds, but it would be useful to have one. Let us use
`gravity-second' to denote a second being used in this way.  The solar
mass in gravity-seconds is
\begin{equation}\label{eq:approx-gsec}
\Mgs_\odot \simeq 4.9\times10^{-6} \sec
\end{equation}
while for the Earth, the value is $15\unit{ps}$.  Again,
Table~\ref{table:conv} gives precise values.  The values of $2\Mgs c$
($3\unit{km}$ for the Sun, $1\unit{cm}$ for the Earth) are more
familiar, but the gravity-second values are not difficult to remember.
Because of the uncertainty in $G$, we do not know precisely how many
kilograms a gravity-second is, but since the constancy of $G$ has been
much better tested than the value has been measured, we do know that a
gravity-second is precisely {\em some} number of kilograms.

One seemingly weird consequence of light-seconds and gravity-seconds
is that density will come out in gravity-seconds per cubic
light-second, which is frequency squared.  To get $\!\kg\metre^{-3}$
we need to divide by $G$, as in
\begin{equation}
\frac{\Mgs}{(4\pi/3) \Rls^3} \times \frac1G \,\up.
\end{equation}
In contexts where particle interactions are of interest, density in
${\rm eV}\metre^{-3}$ may be more useful.  That is easily obtained by
a further factor, as in
\begin{equation}
\frac{\Mgs}{(4\pi/3) \Rls^3} \times \frac{c^2}{Ge} \,\up.
\end{equation}
For gravitational phenomena, however, density as frequency squared is
ideal --- recall that the crossing time in a gravitating system
depends only on the enclosed density, as manifest in, for example, the
destination-independent travel time of 42~minutes on the
``gravity-train'' through the Earth \citep{1966AmJPh..34..703C}.  We
can also write the gravitational constant as
\begin{equation}
1/G = (1.075\unit{hr})^2 \, \unit{g} \unit{cm}^{-3}
\end{equation}
explicitly relating density and time squared.

One thing one musn't do with mass in gravity seconds is to take $\Mgs
c^2$ to get energy!  Instead, to turn gravity-seconds into joules, we
have to multiply by
\begin{equation}\label{eq:planckpower}
\frac{c^5}G = 3.628\times10^{52}\watt \,.
\end{equation}
This constant, sometimes called the Planck power, is the luminosity
scale of merging black holes.

\subsection{Photon flux vs magnitude}

The usual physical measure of brightness in astronomy is the spectral
flux density $f_\nu$.  The relevant SI unit of
$\!\watt\metre^{-2}\Hz^{-1}$ is extremely large, because $1\Hz$ is a
very small spectral range to pack $1\watt\metre^{-2}$, and the
convention in astronomy is to use a jansky, defined as $10^{-26}$ of
the SI unit.  At optical wavelengths, detectors in use nowadays
measure the photon flux in some band
\begin{equation}\label{eq:phfluxold}
\int W_\nu \frac{f_\nu}{h\nu} \, d\nu
\end{equation}
where $W_\nu$ is the throughput of the filter being used.  Many
filters with calibrated transmissions are in use
\citep{2005ARA&A..43..293B}.

Since $h$ is a defined constant in the new SI, let us define
\begin{equation}
\phi_\nu \equiv \frac{f_\nu}h
\end{equation}
which has units of $\hbox{counts}\metre^{-2}\sec^{-1}$.  If we then
change the frequency scale to logarithmic, the photon flux
(\ref{eq:phfluxold}) becomes
\begin{equation}\label{eq:phflux}
\int W_\nu \, \phi_\nu \, d(\ln\nu) \,.
\end{equation}
Applying the condition from \cite{1983ApJ...266..713O} that
$h\phi_\nu=1\unit{erg}\sec^{-1}\unit{cm}^{-2}\unit{Hz}^{-1}=10^{23}\jansky$
corresponds to $\!\ABmag=-48.60$ gives
\begin{equation}
h\phi_\nu = 10^{-22.44-\!\ABmag/2.5} \joule\metre^{-2}
\end{equation}
relating $\phi_\nu$ to $\!\ABmag$.  Dividing by $h$ we have
\begin{equation}\label{eq:approx-phflux}
\phi_\nu = 10^{-\!\ABmag/2.5} \times 5 \times10^{10}
           \metre^{-2}\sec^{-1}
\end{equation}
and the precise coefficient is given in Table~\ref{table:conv}.
Colours are simply the ratio of photon fluxes
$10^{\,\Delta\!\ABmag/2.5}$.

To get the photon flux over a bandpass, we need to take the integral
\eqref{eq:phflux}.  If the spectral density is fairly flat and the
throughput is $\approx1$, the result will be $\sim \phi_\nu \,
\ln(\nu_2/\nu_1)$ where $\nu_2$ and $\nu_1$ are the edges of the band.
If ratio of those is $\simeq1.2$ (which is typical of optical bands)
the photon flux over a band comes to roughly $10^{-\!\ABmag/2.5}
\times1\times10^{10}\metre^{-2}\sec^{-1}$.  In other words, zero AB
magnitude in a typical optical band gives about
$10^{10}\photon\metre^{-2}\sec^{-1}$.

The photon flux $\phi_\nu$ is clearly much more intuitive that the
magnitude scale.  A logarithmic frequency scale also brings some
advantages. First, it does not matter whether the scale is frequency
or wavelength or energy: $\phi_\nu=\phi_\lambda=\phi_E$.  Secondly,
redshift is simply a shift along the spectral scale and does not
change the shape of $\phi_\nu$.

We remark in passing that the SI has three special units relating to
brightness: the candela, the lumen, and the lux.  These units,
however, are designed to quantify the physiological effect of light.
The SI specifies that a watt of monochromatic green light at $540\THz$
is worth 683~lumens, thus defining a lumen.  (A lux is a lumen per
square metre, while a candela is a lumen per steradian.)  Light at
other visible frequencies has fewer lumens per watt, the exact number
being given by a model throughput function for human vision, known as
the luminous efficacy.  Household LED lights deliver about 100 lumens
of white light per watt of electrical power.

\section{Examples}\label{sec:examples}

The preceding sections drew attention to the characteristic style of
the new SI, wherein constants are defined explicitly and units are
defined implicitly thereby, and suggested that the classical
astronomical units for length, mass, and brightness could be replaced
by SI units while leaving metres, kilograms, and watts per hertz
implicit.  We will now discuss examples from different topics in
astrophysics, illustrating how expressing distance in light-seconds,
mass in gravity-seconds, and brightness as $\phi_\nu$ or
$\phi_\lambda$ can be useful in describing diverse astrophysical
processes.

The basic strategy is to work in the SI, while using the freedom given
by the new SI to use any equation of physics to change variables or
introduce new quantities.  Light-seconds, gravity-seconds, and
logarithmic spectral scales are particular instances of this strategy
--- a time variable may be used to represent a length divided by $c$,
and so on.  There is no requirement to use light-seconds for all
lengths, or gravity-seconds for all masses --- we can use whatever
formulation is most convenient, while keeping to the principle that
all dimensional quantities are unambiguously in SI units.  Classical
astronomical units will be converted to SI units at the input stage,
and sometimes classical units will be re-introduced at the output
stage.  What we want to avoid is mixing unit systems inside a
derivation or a computation.  Fundamental constants relating different
SI units (see Table~\ref{table:const}) will be used whenever needed or
useful.  The exception is $G$, which we will use explicitly only where
it is unavoidable, because of its large uncertainty.

For the purposes of this paper, it is not necessary to consider the
most general or most sophisticated form of each process.  The
approximate conversions Eqs.~(\ref{eq:approx-lsec}),
(\ref{eq:approx-gsec}), and (\ref{eq:approx-phflux}) will be useful,
as will some common spherical-cow idealizations.

\subsection{The Sun and the CMB}

Let us compare the nearest and furthest sources of light in astronomy.
For this purpose, we approximate the solar surface by a blackbody at
$T_\odot=5800\kelvin$, and the cosmic microwave background by an ideal
Planckian spectrum with $T_{\rm cmb}=2.725\kelvin$.  The former
approximation is comparatively drastic, but that does not matter for
this example.

The photon flux per logarithmic spectral interval has a simple
interpretation for thermal sources. A blackbody has
\begin{equation}\label{eq:photonspec}
\phi_\lambda = \frac{c}{\lambda^3} \times
               \frac{2\pi}{e^{hc/\lambda kT}-1}
\end{equation}
at its surface.  For a spherical blackbody of radius $r_1$ observed
from distance $r_2$, $\phi_\lambda$ is reduced by a factor
$(r_1/r_2)^2$.  The $\phi_\lambda$ spectrum peaks at $5.100\,{\rm
  mm}\times T^{-1}\kelvin$.  At this peak value, the second factor in
Eq.~\eqref{eq:photonspec} is of order unity (actually $0.40$).  This
makes it useful to imagine the $c/\lambda^3$ as a packed column of
ball-like photons whose diameter is the peak wavelength, moving at the
speed of light.

\begin{figure}
\includegraphics[width=\hsize]{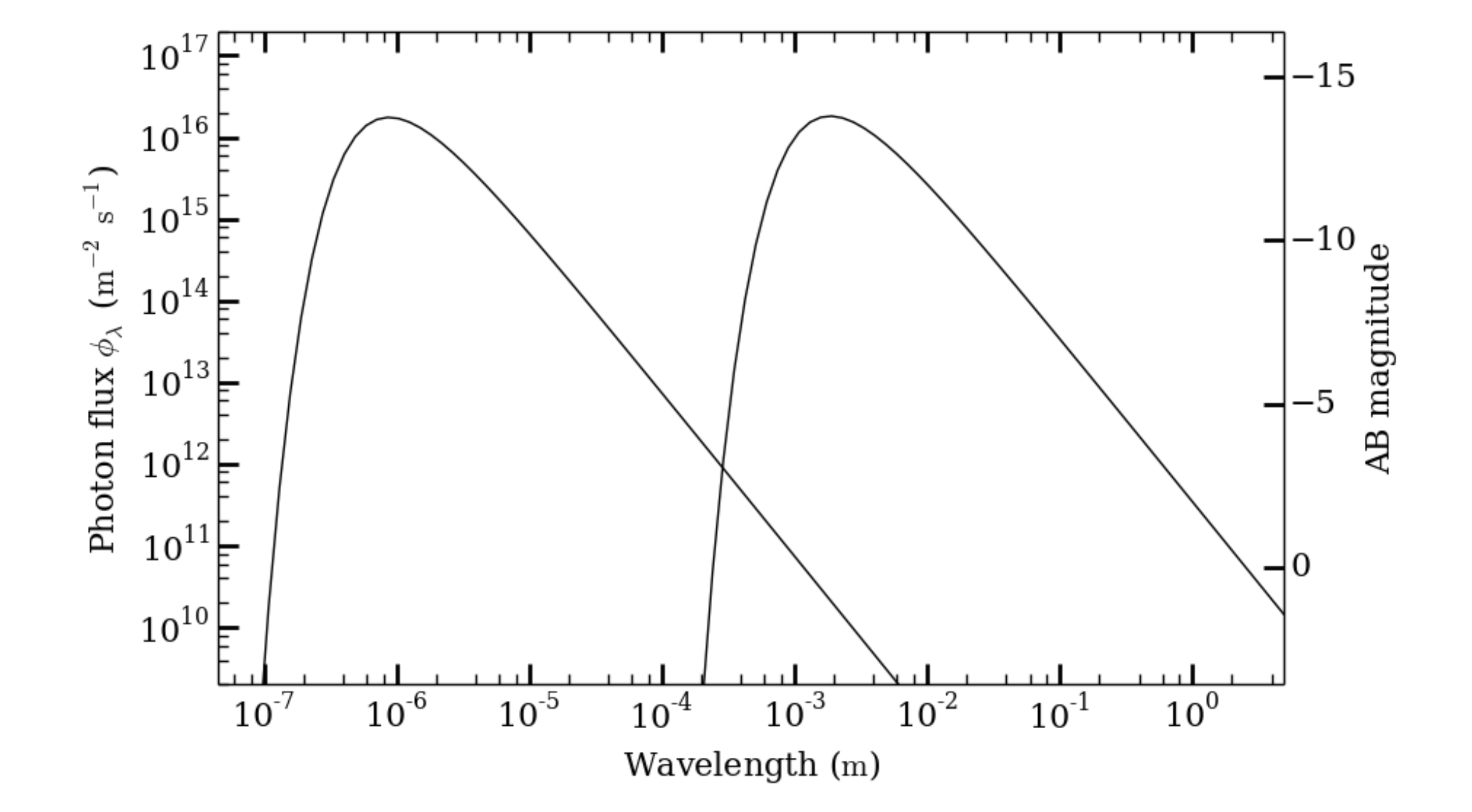}
\caption{Approximate photon spectrum of the Sun, if seen from about
  $450\au$, and the all-sky spectrum from the CMB, shown as
  $\hbox{photons}\metre^{-2}\sec^{-1}$ and AB
  magnitudes.\label{fig:mag}}
\end{figure}

From the photon spectrum \eqref{eq:photonspec} it is clear that the
photon flux will be $\propto T^3$.  Thus, at the solar surface, the Sun
has a photon flux $(T_\odot/T_{\rm cmb})^3\simeq10^{10}$ times that of
the CMB.  From a distance of $(T_\odot/T_{\rm
  cmb})^{3/2}\,R_\odot\simeq450\unit{au}$ the photon fluxes become
equal.  Figure~\ref{fig:mag} illustrates.  The corresponding peaks are
at $1.871\unit{mm}$ (or $160.2\unit{GHz}$) for the CMB and
$880\unit{nm}$ for the Sun.  The latter may seem wrong, because the
solar spectrum is well-known to peak in the middle of the visible
range --- but that is only because the quantity usually plotted for
the solar spectrum is the energy against wavelength, or
$hc\lambda^{-2}\phi_\lambda$.  The AB magnitude of the Sun is indeed
brightest in filters near $900\unit{nm}$ \citep[see
  e.g.,][]{2018ApJS..236...47W}.

\subsection{The Oort and Hubble parameters}

Classical astronomical units often appear mixed with SI units.  Thus,
we may have an expression of the type
\begin{equation}\label{eq:velgrad}
\frac{\partial v_i}{\partial x_j}
\end{equation}
where $v_i$ is a velocity field (mean velocity at location $x_i$),
with $v_i$ measured in $\!\unit{km}\sec^{-1}$ and $x_i$ in
$\!\unit{pc}$.  Dimensionally, such expressions are inverse times, and
it is useful to remember
\begin{equation}\label{eq:approx-vgrad}
1\metre\sec^{-1}\unit{pc}^{-1} = 1\unit{km}\sec^{-1}\unit{kpc}^{-1}
\approx (1\unit{Gyr})^{-1}
\end{equation}
while more digits are given in Table~\ref{table:conv}.  The divergence
(trace of Eq.~\ref{eq:velgrad}) of the velocity field of galaxies is
the Hubble constant.  The (generalized) Oort constants are analogous
quantities for the stellar velocities in the solar neighbourhood.  Of
course, none of these are really constants, just values at our
location or time, so `parameter' may also be used.  Conversions of the
type $100\unit{km}\sec^{-1}\unit{Mpc}^{-1}\simeq(9.78\unit{Gyr})^{-1}$
are often used in both contexts.

The classical Oort parameters $A$ and $B$, defined as
\begin{equation}\label{eq:OortAB}
\begin{aligned}
 2A &= \frac{v_\phi}r - \frac{\partial v_\phi}{\partial r} \\
-2B &= \frac{v_\phi}r + \frac{\partial v_\phi}{\partial r}
\end{aligned}
\end{equation}
in cylindrical coordinates, describe the differential rotation of the
Galaxy.  Solid-body rotation would give $A=0$, whereas a flat rotation
curve gives $A+B=0$.  Recent measurements
\citep[e.g.,][]{2019ApJ...872..205L} give
$A-B\simeq28\metre\sec^{-1}\pc^{-1}$ and all the other components are
an order of magnitude smaller.  Expressing this in SI units as
$A-B\simeq9\times10^{-16}\sec^{-1}$ does not seem much of an
improvement.  If we note, however, that $A-B$ is an angular velocity,
then $A-B=v_\phi/r \simeq 2\pi/(220\unit{Myr})$ is simple.

In the case of the Hubble parameter, it is easy to work with the
Hubble time
\begin{equation}\label{eq:Hinv}
H_0^{-1} = (4.4\pm0.2) \times 10^{17} \sec
         = (14\pm0.5) \Gyr
\end{equation}
which, as well as being the reciprocal expansion rate, is also
approximately the time since the Big Bang.  This already applied in
the old SI.\footnote{For example, \cite{1962ApJ...136..319S} in his
  prediction of redshift drift \citep[which has yet to be observed,
    but see][]{2018EPJC...78...11L} gives $H_0^{-1}=13\Gyr$ as an
  illustrative value, and does not bother with
  $\!\unit{km}\sec^{-1}\unit{Mpc}^{-1}$ at all.}  The new SI, however,
encourages further useful formulations.  Recall that $3/(8\pi G)\times
H_0^2$ is the critical density of the Universe.  The result in
$\kg\metre^{-3}$ is not very intuitive, but with the help of the SI
constants we can also write
\begin{equation}
\frac3{8\pi} \frac{H_0^2}G \times \frac{c^2}e \simeq 5 \unit{GeV} \metre^{-3}
\end{equation}
or roughly 5 atomic mass units per cubic metre.  (Of course, only a
fraction $\Omega_b\simeq0.05$ of this will be in baryonic matter.)  If
the Hubble time is expressed in seconds, no astro-specific
unit-conversion factors are needed, just constants in SI units.
Furthermore, the density expression $3/(8\pi)\times H_0^2$ in
gravity-seconds per cubic light-seconds can also be of use.  Dividing
by $\Mgs_\odot$ gives the density in solar masses per cubic light
second.  We can then calculate a notional distance
\begin{equation}\label{eq:barydens}
\left( \frac3{8\pi} \frac{\Omega_b\,H_0^2}{\Mgs_\odot} \right)^{-1/3}
\sim c\times 5\times10^{10}\sec
\end{equation}
meaning that if all the baryonic matter in the Universe were in
Sun-like stars, these would be on average roughly 500~pc apart.

\subsection{The Milky Way and Andromeda}\label{subsec:timingarg}

We saw above that classical astronomical units can leave us having to
disentangle a mixture of pc, $\!\unit{km}\sec^{-1}$, and $M_\odot$,
with $G$ in SI or cgs units thrown in for good measure.  Another
example where this happens is the Local Group timing argument, in
which the Galaxy and M31 appear as an unusual but very interesting
binary system.

The basic observational facts are that the two galaxies are some
$800\unit{kpc}$ apart and are approaching each other at roughly
$120\unit{km}\sec^{-1}$.  Using this information, one can estimate the
combined mass of these two galaxies by computing how much mass would
be needed to have countered the expansion of the Universe in the Local
Group in a Hubble time.  The idea goes back to
\cite{1959ApJ...130..705K} and has been developed further by many
researchers \citep[e.g.,][]{2016MNRAS.459.2237B}.  The inferred mass
provides a simple and robust estimate of the dark matter in the Local
Group.

Converting the distance to light seconds, we have
\begin{equation}\label{eq:lgparams}
\lgdist \simeq 8\times10^{13}\sec \qquad
d\lgdist/dt \simeq -4\times10^{-4}
\end{equation}
using the notation from Eq.~(\ref{eq:barls}) to denote distance in
light-seconds.  Dividing distance by speed gives a formal time of
$2\times10^{17}\sec$, which is of the same order as the Hubble time.
This suggests modelling the system as a radial two-body orbit that
started to move out at the Big Bang, and has since turned around and
is now approaching a collision.  To do so, let us introduce some
notation for binaries in general.

Consider two masses, $\Mgs_1,\Mgs_2$ in gravity seconds, 
in a two-body orbit, with
\begin{equation}
\Mgs \equiv \Mgs_1+\Mgs_2 \qquad \eta \equiv \frac{\Mgs_1\Mgs_2}{\Mgs^2}
\end{equation}
being the total mass and the symmetric mass ratio.  Let $\als$ be the
orbital semi-major axis in light-seconds, and let us use the
expression
\begin{equation}\label{eq:roemer}
\als = \frac{\Mgs}{\vc^2}
\end{equation}
to define a dimensionless constant $\vc$.  For a circular orbit $\vc$
is clearly the orbital speed in light units.  For general bound
orbits, $\vc$ can be worked out using the virial theorem as the
orbit-averaged rms speed
\begin{equation}
\vc = \frac{\sqrt{\langle v^2\rangle}}c
\end{equation}
in light units.  The orbital period can be expressed as 
\begin{equation}\label{eq:Pdef}
P = 2\pi \frac{\Mgs}{\vc^3}
\end{equation}
thus relating mass in gravity-seconds to an observable time.
Expressing $\als$ in Eq.~\eqref{eq:roemer} as a time is not simply
formal either --- in pulsar binaries \citep[see
  e.g.,][]{2012hpa..book.....L} the light crossing time is the
observable size of the orbit (because no resolved image of the system
is observed), and is known as the Roemer time delay after the
17th-century measurement of the light-travel time across the solar
system.  The Earth's orbit provides a nice illustration of
Eqs.~\eqref{eq:roemer} and \eqref{eq:Pdef}.  Since
$\Mgs_\odot=5\times10^{-6}\sec$ and $\als=500\sec$ we have
$\vc=10^{-4}$ (or $30\unit{km}\sec^{-1}$).  For the orbital period we
recover the well-known mnemonic that a year is $\pi\times10^7\sec$.

Turning now to the Galaxy and Andromeda, let us consider these as
being on a radial two-body orbit.  Such a system has a well-known
solution, which in our notation is
\begin{equation}
\begin{aligned}
t           &= \Mgs\vc^{-3} (\psi - \sin\psi) \\
\lgdist     &= \Mgs\vc^{-2} (1 - \cos\psi) \\
d\lgdist/dt &= \vc \sin\psi/(1-\cos\psi) 
\end{aligned}
\end{equation}
with a formal independent variable $\psi$ serving to give the time
dependence implicitly.  To find the current value of $\psi$, we
consider the dimensionless product
\begin{equation}\label{eq:psiexpr}
\frac t{\lgdist} \times \frac{d\lgdist}{dt} =
\frac{\sin\psi \, (\psi - \sin\psi)}{(1-\cos\psi)^2}
\end{equation}
and in it we put $t=4\times10^{17}\sec$ (about a Hubble time) and the
distance and velocity values from Eq.~\eqref{eq:lgparams} The value of
the expression \eqref{eq:psiexpr} comes to $-2$.  A numerical solution
for $\psi$ yields $\psi\simeq4.2$.  With $\psi$ determined, it is easy
to solve for $\vc$ and $\Mgs$.  The result is
$\Mgs=2.5\times10^7\sec$.  Recalling the conversion
(\ref{eq:approx-gsec}) gives a mass of $5\times10^{12}\Msun$, which is
more than an order of magnitude above the stellar mass in the Local
Group.

In conventional astronomical units, the above steps would be basically
the same.  But because we converted the input to SI at the start, and
converted the inferred mass from gravity-seconds to $M_\odot$ at the
very end, the calculation was much easier. A computer is useful when
solving for $\psi$, but the rest of the arithmetic is trivial.

\subsection{Time dilation in orbiting clocks}\label{subsec:timedil}

When the SI was first instituted in 1960, the second was defined from
astronomy, as a fraction of a mean solar day.  The Caesium-clock
standard changed that a few years later.  Still, astronomy has not
entirely ceded time measurement to atomic physics, because some
applications require time dilation to be taken into account.  The TCB
(temps coordonn\'ee barycentrique) is defined as the time kept by
clocks moving with the solar-system barycenter but outside all
gravitational fields \citep{2001A&A...367.1070B}.

It is well known that global navigation satellites have to correct for
relativistic time dilation. A detailed treatment can be found in
\cite{2003LRR.....6....1A} but a simple estimate makes a nice
illustration of gravity-seconds, as follows.  Let us consider a
two-body orbital system as in subsection~\ref{subsec:timingarg} above,
except that $\eta\rightarrow0$ since the satellite mass is negligible,
and $\Mgs$ is simply the Earth's mass.  Let us write $\Rls$ for the
radius of the Earth in light seconds.  A clock at the surface of the
Earth runs slower than TCB by a fraction $\Mgs/\Rls$, neglecting the
spin of the Earth.  From weak-field relativity, a clock in a circular
orbit of radius $\als$ (in light-seconds) runs slower than TCB by
$\frac32 \Mgs/\als$.  The difference of these two time dilations is
observable.  Multiplying by the orbital period \eqref{eq:Pdef} and
rearranging, we have a delay of
\begin{equation}\label{eq:GPSdelay}
2\pi \left( \frac32 - \frac{\als}{\Rls} \right) \frac{\Mgs}{\vc}
\end{equation}
per orbit.  We can estimate its value from easily-remembered
quantities, as follows: (i)~the circumference of the Earth is
$40\,000\rm\,km$, which gives $\Rls$, (ii)~using $g/c=\Mgs/\Rls^2$ and
putting $g\simeq9.8\metre\sec^{-2}$ gives $\Mgs=15\unit{ps}$
[Table~\ref{table:conv} gives the precise value.]  (iii)~equating
$2\pi\Mgs/\vc^3$ to the orbital period of $12\unit{hr}$ for GPS
satellites gives $\vc$, (iv)~using $\vc^2=\Mgs/\als$ gives $\als$.
Doing the arithmetic, we find a delay of $38.5\microsec$ per day.

GPS clocks are also in orbit around the Sun.  Hence, though they tick
faster than terrestrial clocks, satellite clocks run slower than a
notional TCB clock.  To find the time delay over one orbit (that is,
one year), we use the first term in Eq.~\eqref{eq:GPSdelay} and
substitute the solar mass and the Earth's orbital speed.  The latter,
we have already seen, is yet another conveniently round number
$\vc=10^{-4}$.  The result is $0.5\sec$ per year.  The solar mass
parameter (cf.\ Eq.~\ref{eq:GMsun}) is
$G\Msun=1.32712442099(10)\times10^{20}\metre^3\sec^{-2}$ in TCB but
measurably different with respect to terrestrial time
\citep{Luzum2011}.

For eccentric orbits, the time dilation is more complicated in detail,
but of the same order.  A good example is the star S2/S0-2, which
orbits the black hole at the center of the Milky Way in a highly
eccentric orbit.  The spectral features of the star amount to a
natural clock, and its orbital time dilation has recently been
measured \citep{2018A&A...615L..15G,2019Sci...365..664D}.  Further
observable times of the form $\Mgs\vc^n$ (with positive $n$) can
emerge from higher-order relativistic effects
\citep{2014MNRAS.444.3780A} but have not been measured yet.

\subsection{Shapiro and Refsdal delays}

As well as the various $\Mgs\vc^{-n}$ in seconds, there is a
measurable time that is simply $\Mgs$ times a numerical factor.  As
one might guess from the absence of $\vc$, it involves light.

A light ray, that on its way between source and observer has flown
past a mass $\Mgs$, experiences a time delay
\begin{equation}\label{eq:lightdelay}
-2\Mgs\ln(1-\cos\theta)
\end{equation}
where $\theta$ is the angle on the observer's sky between the mass and
the incoming ray.  The logarithm in \eqref{eq:lightdelay} will be
negative, assuming $\theta$ is not too large, and hence the whole
expression will be positive.  This delay is well known in ranging
experiments in the solar system, and in pulsar timing, and is known as
the Shapiro delay after the prediction by \cite{1964PhRvL..13..789S}.

Another manifestation of the same phenomenon, predicted by
\cite{1964MNRAS.128..295R} by very different arguments, appears in
gravitational lensing.  In the regime of lensing, $\theta$ is small.
Taking the small-$\theta$ limit of \eqref{eq:lightdelay} gives
$-2\Mgs(2\ln\theta-\ln2)$.  But at small $\theta$, the deflection
$\Delta\theta$ of the light ray is important, as it contributes a
further time delay.  The time delay depends on $\theta$ and
$\Delta\theta$ as
\begin{equation}\label{eq:arriv}
t(\theta) = \Dls \, \Delta\theta^2 - 4\Mgs\ln\theta
\end{equation}
where $\Dls$ is an effective distance (in light-seconds) depending on
the distances to the mass and the light source.  Light then follows
Fermat's principle and chooses $\theta$ and $\Delta\theta$ so as to
make the total light travel time extremal
\citep[e.g.,][]{1986ApJ...310..568B}.  There can be more than one
extremal light path, giving multiple images with different light
travel times.  A good example is Supernova Refsdal observed to appear
at displaced locations with time delays \citep{2016ApJ...819L...8K}.
In general, a gravitational lens will not be a single mass but an
extended mass distribution.  Accordingly, the last term in
Eq.~\eqref{eq:arriv} has to be replaced by an integral over the mass
density.  The observable time delay between different lensed images
then depends on details of the how the mass is distributed \citep[see
  e.g.,][]{2015PASJ...67...21M}.

If a mass is at a cosmological distance (say at redshift $z$), the
Shapiro or Refsdal delay as measured by an observer will be
time-dilated by $(1+z)$, just like any other time interval at that
redshift.  That is, mass in gravity-seconds gets redshifted.

\def\cpm{\kappa}

\subsection{Gravitational-wave inspiral}

In general relativity it is common to put $G=c=1$, leaving units and
dimensions implicit \citep[geometrized units --- see Appendix F
  in][]{1984ucp..book.....W}.  Light seconds and gravity seconds are
similar, but with explicit variable changes to keep track of units and
dimensions, and using these we can express gravitational-wave inspiral
very concisely, while also keeping the comparison with observations
simple.

Let us again consider a binary with parameters $\Mgs,\eta,$ and $\vc$.
For this system let us write
\begin{equation}\label{eq:omega}
\omega = \vc^3 / \Mgs
\end{equation}
for the angular orbital frequency, and
\begin{equation}\label{eq:Edef}
\Egs = -{\textstyle\frac12} \eta \Mgs \vc^2
\end{equation}
for the orbital energy in gravity-seconds.

Such a binary will produce gravitational waves of angular frequency
$2\omega$, and the strain components at distance $\Dls$ will be
\begin{equation}\label{eq:hgw}
\hgw \sim \frac{\Egs}{\Dls}
\end{equation}
with numerical coefficients of order unity, depending on orientation.
It may be convenient to think of the strain as a kind of gravitational
potential whose source is not the mass but the orbital energy.  If we
further associate the wave with an energy density
$\propto\omega^2\hgw^2$ \citep[cf.][]{2017AmJPh..85..676M} and moving
with a speed of light, it follows that energy will be emitted at a
rate $\propto\omega^2\Egs^2$.  Writing the proportionality constant as
$\cpm$, we have
\begin{equation}\label{eq:gwrad}
\frac{d\Egs}{dt} = -\cpm \, \omega^2 {\Egs}^2 \,.
\end{equation}
This paragraph is not a derivation, just a plausibility argument.
Nevertheless, the expression \eqref{eq:gwrad} with $\cpm=128/5$ turns
out to be the correct leading-order relativistic result for circular
binaries \citep{1963PhRv..131..435P}.  Eccentric orbits and higher
order modify the numerical factor, but do not change the units needed.
Notice that the power output \eqref{eq:gwrad} is in gravity-seconds
per second.  To convert to watts, we need to multiply by $c^5/G$ (as
noted at Eq.~\ref{eq:planckpower}).

Low-mass gravitationally radiating binaries can be just as luminous as
supermassive binaries, but the latter take longer.  To see this, we
eliminate $\Egs$ and $\omega$ from the formula \eqref{eq:gwrad} and
rearrange to get
\begin{equation}\label{eq:vdot}
\frac{d\vc}{dt} =
{\textstyle\frac14} \cpm \eta \times \frac{\vc^9}{\Mgs}
\end{equation}
giving the increasing speed of the inspiralling system.  Integrating
to get the time left before $\vc=1$ (when the system would merge), we
get
\begin{equation}\label{eq:tinsp}
   T_{\rm insp} = \frac1{2\cpm\eta} \, \frac{\Mgs}{\vc^8}
\end{equation}
for the inspiral time.  Here we have yet another time scale of the
form $\Mgs/\vc^n$.

The parameters $\Mgs,\eta,$ and $\vc$ are, in general, not directly
observable for inspiralling binaries.  The observables are the two
components (polarizations) of the strain, the wave frequency
$2\omega$, and its derivative, known as the chirp.  In terms of there
we can write
\begin{equation}\label{eq:chirp}
\frac1\omega \frac{d\omega}{dt} =
{\textstyle\frac38} T_{\rm insp}^{-1} = 
{\textstyle\frac32} \cpm \omega^2 \Egs .
\end{equation}
Since the left expression is observable, the other two expressions
become measurable.  That $T_{\rm insp}^{-1}$ can be inferred from the
frequency and chirp is not surprising.  That $\Egs$ is measurable is
remarkable, and has important consequences.  Writing $\Egs$ as
$\frac12\eta\Mgs^{5/3}\omega^{2/3}$, we see that the combination
$\eta^{3/5}\Mgs$ is also measurable.  It is known as the chirp mass.
Still more interesting is that gravitational-wave binaries can serve
as ``standard sirens'' enabling distance measurements, as first noted
by \cite{1986Natur.323..310S}.  To see how, recall from
Eq.~\eqref{eq:hgw} that $\Egs$ relates strain and distance.  The
coefficients in the equation can be determined from the measurable
polarization of the gravitational wave.  This makes $\Dls$ measurable
too, and from it $H_0$ as well if the redshift is measured.

Regarding redshifts, it is important to note that a redshift $z$
dilates $t$ in the observer frame by $1+z$.  As a result, observed
frequencies are slowed down, and the inferred chirp mass is dilated,
by the same factor.  The speed $\vc$ is not affected.  To make the
strain in Eq.~\eqref{eq:chirp} come out right, we need to dilate
$\Dls$ by $1+z$ as well (i.e., use the luminosity distance).  Again,
as was the case in lensing, we see mass in gravity-seconds at
cosmological distances getting redshifted.

GW170817 \citep{PhysRevLett.119.161101} was a particularly interesting
gravitational-wave source, providing a wealth of
observations in addition to the inspiral, and taken together these
enabled a reconstruction of the system as two neutron stars at $\simeq
40\unit{Mpc}$, which corresponds to
$$ \Mgs \simeq 1.5\times10^{-5}\sec \qquad
   \eta \simeq {\textstyle\frac14} \qquad
   \Dls \simeq 4\times10^{15}\sec .
$$
The level of strain at merger would be $\Mgs/\Dls\sim10^{-21}$.  If we
assume $\vc=0.12$ at the start of the detected event,
Eq.~\eqref{eq:omega} gives an initial orbital frequency of
$\simeq20\Hz$ (wave frequency of $\simeq40\Hz$), while
Eq.~\eqref{eq:tinsp} implies $T_{\rm insp}\simeq 30\sec$, reproducing
the observed values.

Galactic binary pulsars are younger systems like GW170817.  The
gravitational-wave inspiral of these systems has been known since the
first measurement by \cite{1979Natur.277..437T}, but detecting the
gravitational-wave strain from these low-frequency sources is not
feasible yet.  Interestingly, the $\hgw$ from Galactic binary pulsars
is comparable to that from GW170817.  To see this, we can simply scale
$$ \vc\rightarrow10^{-2}\vc, \qquad \Dls\rightarrow10^{-4}\Dls $$
which leaves $\hgw$ the same.  The distance changes to $4\unit{kpc}$,
a $20\Hz$ orbital frequency ($\propto\vc^3$) changes to a half-day
orbital period, and a $30\sec$ inspiral time ($\propto\vc^{-8}$)
changes to $10^{10}\unit{yr}$.  These values are typical of Galactic
binary pulsars \citep[see][]{2008LRR....11....8L}.

\subsection{Eddington luminosity and the M87 black hole}

In the preceding examples, matter only contributed a gravitational
field, and hence mass always appeared multiplied by the gravitational
constant, which we were able to absorb inside $\Mgs$, thus eliminating
the uncertainty in $G$.  If matter contributes in other ways too (such
as producing gas pressure), mass will appear as both $M$ and $GM$.
This will make the uncertainty in $G$ unavoidable --- and also offer a
way to measure $G$, as in \cite{2005MNRAS.356..587C}.

An exceptional but important situation, in that mass appears only as
$GM$ even though non-gravitational processes are involved, is
Eddington luminosity.  In this one has a spherical mass $M$ of ionized
gas, and some energy source which gives it a luminosity $L$, and the
radiation pressure from the latter balances the self-gravity.  At any
radius $r$ inside the sphere, we have
\begin{equation}
\frac{GMm_p}{r^2} =
\frac{L/c}{4\pi r^2} \times \frac2{3\pi} \left(\frac{\alpha h}{m_ec}\right)^2 
\end{equation}
where $m_p$ and $m_e$ are the masses of the proton and electron, and
$\alpha$ is the fine-structure constant.  On the left of this equation
we have the weight of an ion, and on the right we have the outward
momentum flux times the Thomson cross-section.  Radiation pressure
acts on the electrons, but the force is transmitted electrostatically
to the ions.  Expressing the particle masses as equivalent frequencies
$\nue=m_ec^2/h,\nup=m_pc^2/h$ and rearranging, we get the Eddington
luminosity
\begin{equation}
L = 6\pi^2 \, h\nup \, (\nue/\alpha)^2 \Mgs
\end{equation}
written with mass in gravity-seconds.

Although originally developed for massive stars, the Eddington
luminosity is nowadays also often applied to accreting black holes to
estimate the maximum possible luminosity.  Accretion by black holes is
not a spherical process, so applying the formula to black holes gives
a rough estimate at best, but is nonetheless interesting.  Let us
accordingly approximate an accreting black hole as a blackbody sphere
at temperature $T$.  It will radiate at
$2\pi^5c/15\times(kT)^4/(hc)^3$ per unit area.  For the radius of the
sphere, we take the radius of the innermost stable orbit, which is
$6\Mgs$ (or $6\Mgs c$ in length units) for a non-spinning black hole.
Equating the total luminosity to the Eddington luminosity we can
define an effective temperature
\begin{equation}\label{eq:bhtemp}
T = \frac h{2\pi k} \left(\frac{5\nue^2\nup}{\alpha^2\Mgs}\right)^{1/4}
\end{equation}
for the accreting system.  The formula for this ``Eddington
temperature'' seems strangely reminiscent of the much much colder
Hawking temperature $h/(4k\Mgs)$.

\cite{2019ApJ...875L...1E} present an interferometric image of the
silhouette of the supermassive black hole in M87 at a resolution of
$20\,\mu{\rm as}$ or $10^{-10}\unit{radians}$.  The resolution is as
expected for mm-wavelengths with a baseline of $\sim10^4\unit{km}$.
The distance to M87 being $\simeq 20\unit{Mpc}$, the resolved size
comes to $2\times10^{-3}\pc$, which is like $2\times10^5\sec$ or two
light days.  The inferred mass is about an order of magnitude smaller
than this scale: $6\times10^9\Msun$ or $3\times10^4$ gravity-seconds.
Plugging the mass in the formula \eqref{eq:bhtemp} gives
$T\simeq8\times10^4\kelvin$.  Taken as an upper limit, this value is
very reasonable, since a continuum peak around $100\unit{nm}$,
corresponding to an effective temperature of
$T\approx3\times10^4\kelvin$, is typical of quasars
\citep{1991ApJ...373..465F,2001AJ....122..549V}.  The M87 black-hole
system itself would have a much lower effective temperature, because
it is accreting only weakly now.  The measured brightness temperature
at mm wavelengths is, however, much higher.  This tells us that the mm
radiation cannot be thermal and must be predominantly reprocessed.

\section{Discussion}

The recent reforms of the SI have made the formerly unexciting subject
of units scientifically novel.  Astronomers, however, have always been
unwilling to adopt SI units.

The persistence of classical and other pre-SI units in astrophysics
actually has an interesting scientific reason, namely the difficulty
of calibrating astronomical observables against the laboratory
standards on which the SI and its predecessors are based.  The
calibration problems are mostly solved now, but one very important
problem remains: the uncertainty in $G$, which makes the kilogram
unusable in some precision applications.  Without the kilogram, any
proposal to change to SI units becomes a non-starter.  At most, one
sees arguments for a partial change to SI units
\citep[see][]{2011usua.book.....D}.  And so it is that every new
research student in astronomy, having mastered basic physics with SI
units, is confronted with magnitudes, parsecs and solar masses, as
well as pre-SI decimal metric units like \AA, ergs, and gauss.
Expressions mixing different unit systems are especially painful, and
make it difficult to catch errors.

The new SI, by giving physical constants the central role, encourages
reformulation of observables by inserting factors of $c,h,$ and so on.
With this freedom, the classical units of length, brightness, and mass
can be usefully replaced by SI units having different dimensions.
\begin{itemize}
\item The au and pc are easily replaceable by light-seconds, and both
  conversions are close to round numbers ($500\sec$ and $10^8\sec$).
\item AB magnitude is equivalent to the photon flux per logarithmic
  spectral interval, which is much easier to understand and work with.
  For typical optical bands, zero magnitude $\simeq
  10^{10}\unit{photons}\,\metre^{-2}\sec^{-1}$.
\item Measuring astronomical mass in gravity-seconds may seem
  contrived at first, but it is really a simple variant of the mass
  parameter used in solar-system dynamics, and can help gain new
  insight into diverse astrophysical processes.  Especially nice is
  how some very different observable time scales in orbital systems
  are set by the mass in gravity-seconds and the dimensionless speed:
  the classical Roemer delay is $\sim\Mgs/\vc^2$ and the orbital
  period is $2\pi\Mgs/\vc^3$, in general relativity the time dilation
  per orbit is $\sim\Mgs/\vc$, the gravitational-wave inspiral time is
  $\sim\Mgs/\vc^8$ and the Shapiro and Refsdal delays are $\sim\Mgs$.
\end{itemize}
The light-second and gravity-second are not to be considered as new
(and therefore non-SI) units with dimensions of length and mass.  They
are simply the second being used to measure a length times $c^{-1}$,
or measure a mass times $G/c^3$.

An indirect benefit of the classical astronomical units is that
working astronomers are quite used to converting between different
unit systems.  Theoretical calculations of idealized systems may be
done in geometrized units or even Planckian units
\citep[e.g.,][]{magicenv}, and compute-intensive work often favours
internal units to improve numerical performance.  All of these require
unit conversion at input and output stages, but provided unit
conversion is a minor overhead, it does not cause problems.  Hence, SI
replacements for the classical astronomical units can simply be
incrementally introduced by early adopters, without requiring any
formal policy changes.

All that said, which astronomer does not love their parsecs and
magnitudes?  Moreover, there are parameters that have a parsec inside
their definitions: absolute magnitude and the conventional
normalization of the cosmological power spectrum $\sigma_8$.  Now,
there is an interesting social phenomenon that sometimes the word for
an archaic unit survives, but changes its meaning to a round number of
the new unit.  For example, contemporary German usage has rounded up a
`Pfund' from a pound to $500\unit{g}$, while in South Asia a `tola'
has been rounded down to $10\unit{g}$.  One can imagine the same for
the classical astronomical units.
\begin{itemize}
\item A rounded solar-mass unit as $\Mgs_\odot=5\times10^{-6}\sec$ has
  already been used in this paper.
\item Rounded parsecs of exactly $10^8$ light-seconds (which is just
  over $\pi$ light-years) would be 3\% smaller than parsecs.  For
  $\sigma_8$ in cosmology, the power-spectrum would get averaged over
  a volume about 10\% smaller, and the change in that average may be
  insignificant.  A rounded au of 500~light-seconds would also be
  useful.
\item Zero magnitude could be conveniently rounded to $10^{10}\hbox{
  photons } \metre^{-2}\sec^{-1}$, with reference to a broad band
  whose width is 20\% of its median.
\end{itemize}
Rounding the classical astronomical units in this way would be
harmless for most applications.

\bigskip

Thanks to P.R.~Capelo, J.~Magorrian, A.~Saha, R.~Sch\"onrich,
L.L.R.~Williams, and the referee for comments on earlier versions.

\def\aj{AJ}
\def\apj{ApJ}
\def\apjl{ApJL}
\def\apjs{ApJS}
\def\aap{A\&A}
\def\araa{ARA\&A}
\def\mnras{MNRAS}
\def\nat{Nature}
\def\pasj{PASJ}
\def\prd{Physical Review D}
\def\jgr{JGR}
\def\sovast{Sov. Astr.}
\def\rsquo{'}

\bibliography{ms.bbl}

\appendix

\section{Constants and conversion factors}

This Appendix summarizes the various constants and conversion factors
used in this paper.  Table~\ref{table:const} consists of physical
constants relevant to astronomy, while Table~\ref{table:conv} has
conversion factors from the classical astronomical units.

\begin{table}
\centering
\caption{Physical constants in the new SI\label{table:const}}
\smallskip
\begin{tabular}{ccl}
\toprule
$c$            & & $299792458\metre\sec^{-1}$ \\[3pt]
$h$            & & $6.62607015\times10^{-34}\joule\sec$ \\[3pt]
$e$            & & $1.602176634\times10^{-19}\joule\volt^{-1}$ \\[3pt]
$k$            & & $1.380649\times10^{-23}\joule\kelvin^{-1}$ \\[5pt]
$G$            & \kern30pt  & $6.674 3(2)\times10^{-11}
                               \kg^{-1}\metre^3\sec^{-2}$ \\[3pt]
$1/\alpha$     &            & $137.035 999 08(2)$ \\[3pt]
$m_e$          & & \llap{$e/c^2\,\times\,$}$0.510 998 950 0(2)\unit{MeV}$
                   \\[3pt]
$m_p$          & & \llap{$e/c^2\,\times\,$}$0.938 272 088 2(3)\unit{GeV}$ \\
\bottomrule
\end{tabular}
\end{table}

\begin{table}
\centering
\caption{Classical astronomical units in SI terms\label{table:conv}}
\smallskip
\begin{tabular}{ccl}
\toprule
${\rm AB}=0$   &   & $5.4795384 \times 10^{10} \metre^{-2} \sec^{-1}$ \\[5pt]
$\au$          &   & \llap{$c\,\times\;$}$499.00478 \sec$ \\[3pt]
$\pc$          &   & \llap{$c\,\times\;$}$1.0292713 \times 10^8\sec$
                     \\ [5pt]
$\metre\sec^{-1}\unit{pc}^{-1}$
               &   &$3.2407793\times10^{-17} \sec^{-1}$ \\
               &   & \llap{$($}$0.9777922 \unit{Gyr})^{-1}$ \\[5pt]
$\Msun$        &   & \llap{$c^3/G\,\times\;$}$4.9254909 \times 10^{-6} \sec$
                     \\[3pt]
$\Mearth$      & \kern24pt & \llap{$c^3/G\,\times\;$}$1.4793661 \times 10^{-11}
                 \sec$ \\
\bottomrule
\end{tabular}
\end{table}

The first four constants in Table~\ref{table:const} have defined
values in the new SI.  The others are experimentally determined and
hence have uncertainties.  Neither is a complete set, but simply the
subset important in astrophysics.  Taking advantage of equivalences in
the new SI, electric charge is written in joules per volt, and
particle masses in $e/c^2$ times volts.  It is worth mentioning that
the vacuum permeability $\mu_0$ is no longer a defined constant;
instead, permeability and permittivity are given by
\begin{equation}
c\mu_0 = \frac1{c\epsilon_0} = 2\alpha\frac{h}{e^2}
\end{equation}
the latter constant being the vacuum impedance $\simeq377\ohm$.

Table~\ref{table:conv} expresses the classical astronomical units in
terms of SI units.  Note that in each case, some simple change of
variable is involved.  The first four numerical factors are actually
exact numbers (that is, derived from defined constants) but have been
rounded to eight digits in the table.  The mass values are measured
quantities whose current uncertainties are smaller than the eight
digits given here.  As noted in the main text, all the numerical
values are close to some round number and hence easy to remember
approximately.

\end{document}